# Taurus stars membership in the Pleiades open cluster

Tadross, A. L., Hanna, M. A., Awadalla, N. S.

*National Research Institute of Astronomy & Geophysics, NRIAG, 11421 Helwan, Cairo, Egypt*

**ABSTRACT**

In this paper, we study the characteristics and physical properties of the young open cluster Pleiades using Near Infra-Red, JHK pass bands. Our results have been compared with those found in new optical UBV observations. The membership validity of some variable binary stars, which are Located in Taurus constellation, and their relation with Pleiades cluster have been achieved.

## 1. INTRODUCTION

Pleiades and Hyades are the most famous star clusters in Northern Hemisphere which can be seen by the naked eye in the constellation Taurus the bull. The star cluster surrounding Aldebaran (the eye of the bull) is the Hyades. Pleiades (NGC 1432; M45; Melotte 22; Seven Sisters) is more famous than Hyades because it is more compact cluster and easy seen, higher in the sky (~ 10 degrees from Aldebaran), see Fig 1. Pleiades is located at J (2000) $\alpha = 03^h: 47^m: 24^s$; $\delta = +24^o: 07': 12''$; G. long. = $166.642^o$ and G. lat. = $-23.457^o$. The distance to the Pleiades is an important first step in the so-called cosmic distance ladder, a sequence of distance scales for the whole universe. The optical photometry catalogs of Webda and Dais refer to Pleiades as a rich young cluster located at 135-150 pcs away from Earth. In optically observations, Pleiades covers a diameter of about 110 arcmin on the sky; its core and tidal radii are about 33 and 330 arcmins respectively. The cluster contains statistically over 1000 members, excluding unresolved binary stars. It is dominated by young, hot blue stars, which were formed within the last 135-150 million years. The dust that causes faint reflection nebulosity around the brightest stars is the remnants that left over from the very beginning, formation of the cluster; $E_{B-V}$ = 0.03 mag; see Fig 2. The total mass of the cluster is about 800 $\mathcal{M}_\odot$. On the other hand, it contains many brown dwarfs, about 25% of the star cluster, which are objects with less than about 8% of solar mass, i.e. not heavy enough for nuclear fusion reactions to start inside.

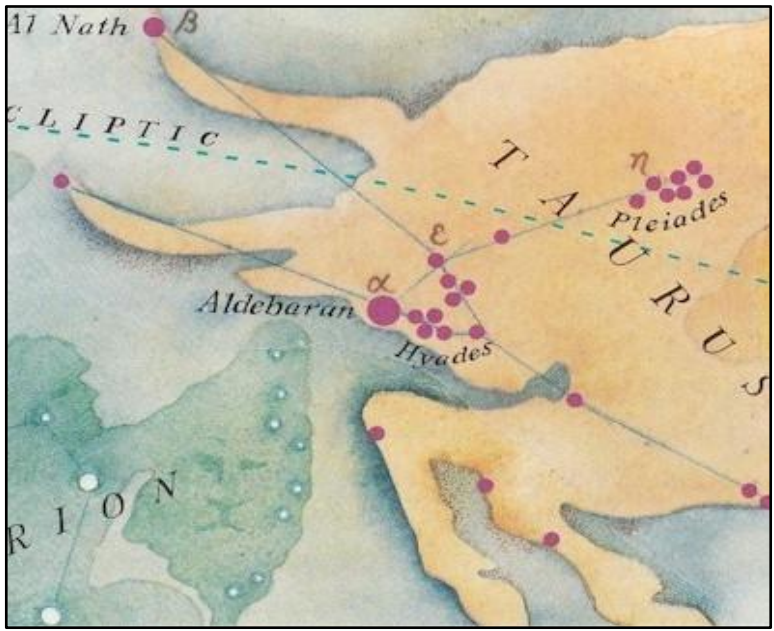

Fig 1: The constellation Taurus.

Tadross, A. L., Hanna, M. A., Awadalla, N. S.

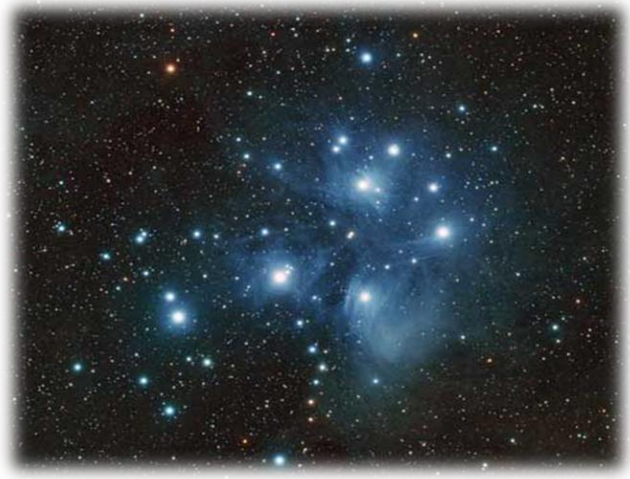

Fig. 2: The open star cluster Pleiades (NGC 1432; M45/ Melotte 22 / Seven Sisters).

North is up, and East on the left.

## 2. DATA EXTRACTION AND FIELD DECONTAMINATION

Data extraction has been performed using the known tool of VizieR[1]. The number of stars in the direction of Pleiades within a radius of 90 arcmin in the range of $5 \leq J \leq 13$ mag is found to be 2655 stars. Usually, field stars contaminate the CMDs of the cluster, particularly at faint magnitudes and red colors. In order to define the intrinsic CMDs of the cluster, we have to compare the CMDs of Pleiades with a nearby control field of the same area as the cluster. A control field is chosen at the same Galactic latitude, but with one degree away from the Galactic longitude of the cluster's center. Comparing the CMDs of the cluster and its control field at a given magnitude and color range; the number of stars in the control field should be subtracted from that of the cluster. In this respect, for more accuracy, our data has been restricted to J, H, K<0.20 mag due to the observational uncertainties. Applying a cutoff of photometric completeness (J<13 mag) to both cluster and control field to avoid what is called over-sampling (Bonatto et al. 2004; Tadross 2008).

---
[1] http://vizier.u-strasbg.fr/viz-bin/VizieR?-source=2MASS

### 3. PLEIADES' DIAMETER ESTIMATION

To determine the cluster's minimum radius, core and tidal radii, the radial surface density of the stars should be constructed first. The tidal radius determination is made possible by the spatial coverage and uniformity of 2MASS photometry, which allows one to obtain reliable data on the projected distribution of stars over extended regions around clusters (Bonatto et al. 2005). We found that the background contribution level corresponds to the average number of stars included in the comparison field sample is 0.93 stars per arcmin$^2$. Applying the empirical profile of King (1962), the cluster's apparent radius turns out to be 70 arcmin, as shown in Fig 3.

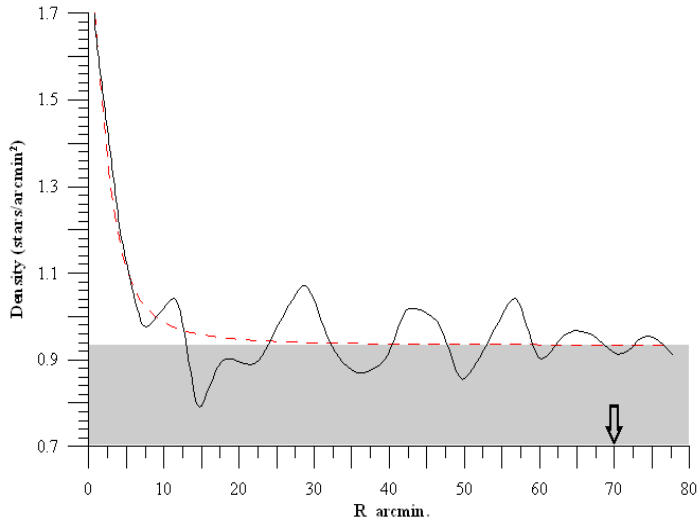

Fig 3: Radial distribution of the surface density of Pleiades (solid curve). The dashed line represents the fitting of King (1962) model. The arrow refers to the radius of the cluster (70 arcmin). The shaded region shows the mean level of the comparison field density, taken to be 0.93 stars per arcmin$^2$.

## 4. COLOR-MAGNITUDE DIAGRAM ANALYSIS

Because the background field of Pleiades is very crowded and the cluster`s data is contaminated, the CMDs of Pleiades can be constructed with the stars inside radii of 1, 2, 3, etc., arcmin from the cluster center. Theoretical Padova isochrones of the solar metallicity (Z=0.019) with J, H, and $K_S$ colors of Bonatto et al. (2004) have been used in fitting to derive the cluster parameters. Simultaneous fittings were attempted on the J ~ (J-H) and $K_S$ ~ (J-$K_S$) diagrams for the inner stars, at which they should be less contamination by the background field. If the number of stars were not enough for an accepted fitting, the next larger area would be included, and so on. In this respect, different isochrones of different ages have been applied on the CMDs of Pleiades, fitting the lower envelope of the points matching the main sequence stars, turn-off point, and red giant positions. The average age, reddening and distance modulus, within a ranging fitting error of about ± 0.10 mags, are determined. The data has been corrected for interstellar reddening using the coefficients ratios $\frac{AJ}{Av}$ = 0.276, $\frac{AH}{Av}$ = 0.176 and $\frac{AKs}{Av}$ =0.118 which were derived from absorption ratios in Schlegel et al. (1998), and Dutra et al. (2002) where $R_v$ = 3.1.

Fig 4 shows the CMDs of the Pleiades, with magnitude completeness limit and the color filter for the stars within the apparent cluster radius of 70 arcmin. The membership criterion adopted here for inclusion of stars in CMDs that, they must be close to the cluster main sequence, deviating by no more than about 0.10 mags. On this basis, the fundamental photometric parameters of the cluster (reddening, apparent distance modulus, and age) can be determined simultaneously.

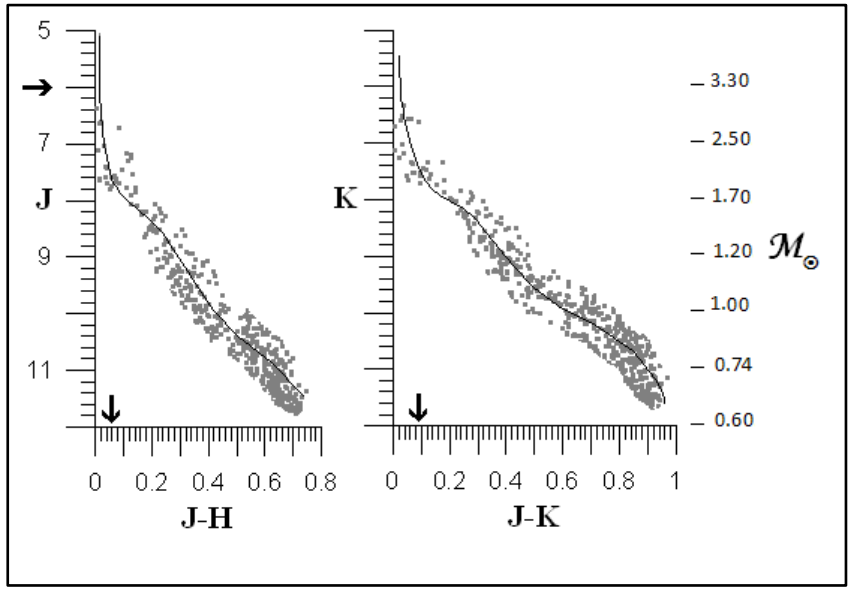

Fig 4: Padova solar isochrone (solid lines) with age of 150 Myr is fitted to the CMDs of Pleiades. Color and magnitude filters have been used in reducing the field star contamination of the cluster. The horizontal and vertical arrows refer to the values of distance modulus, and the color excess in each diagram, on the vertical and horizontal axes respectively. The scale of masses is presented on the right side of the figure, expressed in solar mass.

The overall shape of CMDs is found to be well reproduced with isochrones' of age of 150 Myr old. The apparent distance modulus is taken at 6.00±0.10 mag, accordingly the distance modulus of 5.83±0.10 mag, which corresponds to a distance of about 147±8 pc is most acceptable. On the other hand, the color excess $E_{J-H}$ is found to be 0.06±0.03 mag, which corresponds to $E_{B-V}$ of about 0.20 mags, which is in agreement with Schlegel, et al. (1998).

The number of stars in the region of Pleiades is about 400 stars, with a mean mass of 1.9 $\mathcal{M}_\odot$. Then the total mass of Pleiades is 760 $\mathcal{M}_\odot$. It is noted that the mass estimation for unresolved binaries and low mass stars is always problematic. Jaschek & Gomez (1970) claimed that approximately 50% of the main sequence stars might be hidden. According to this assumption, the total mass of Pleiades can be as large as 1000 $\mathcal{M}_\odot$. Knowing the cluster`s distance in parsecs, the linear diameter can be easily estimated to be 6 pc. Applying the equation of Jeffries et al. (2001), the tidal radius of Pleiades is found to be 15 pc. The present photometric *JHK* work have been compared with previous *UBV* studies, e.g. WEBDA[2] and DIAS[3] catalogs, the results can be seen in Table 1.

## 5. THE VARIABLES IN TAURUS CONSTELLATION

The constellation "Taurus" contains many types of variable stars, the position of those located in the same area of Pleiades have been counted and drawn on the face of the cluster, as shown in Fig 5. About 160 variables are found concentrated around the cluster center. The membership validity of those variables in the cluster area has been tested and expressed on the same CMDs of Pleiades, after applying the color and magnitude filters. 63 of them (~ 16% of the cluster's stars) are found to lie very close to the cluster's CMDs, and so their membership probability to Pleiades is photometrically verified, as shown in Fig 6. Table 2 contains the details of the 63 variable members in Pleiades.

---

[2] *http://obswww.unige.ch/webda*

[3] *http://www.astro.iag.usp.br/_wilton/*

Table 1: Comparisons between the photometric parameters of the present *JHK* work and the previous *UBV* studies

| Parameter | Present work (JHK) | Previous studies (UBV) |
|---|---|---|
| Distance modulus | 6.00±0.10 mag. | 5.97 mag. |
| Distance | 147±8 pc. | 150 Pc. |
| $E_{J-H}$ | 0.06±0.03 mag. | --- |
| $E_{B-V}$ | 0.20 mag. | 0.03 mag. |
| Age | 150 Myr. | 135-150 Myr. |
| Radius | 70 arcmin (6 pc.) | 60 arcmin |
| Tidal radius | 15 pc. | --- |
| Z | -60 pc. | -59.9 pc. |
| $R_{gc}$ | 9 kpc. | --- |
| $X_\odot$ | 134 pc. | --- |
| $Y_\odot$ | 32 pc. | --- |
| Total mass | ~ 1000 $M_\odot$ | ~ 800 $M_\odot$ |

Where Z, $R_{gc}$, $X_\odot$ and $Y_\odot$ are the distance from galactic plane, the distance from the galactic center, and the projected distances on the galactic plane from the Sun, respectively.

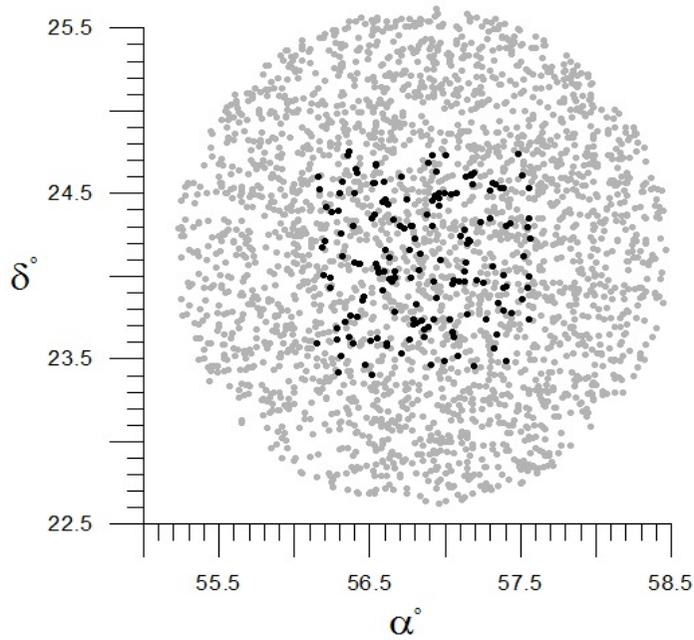

Fig 5: The variable stars that located on the face of Pleiades.

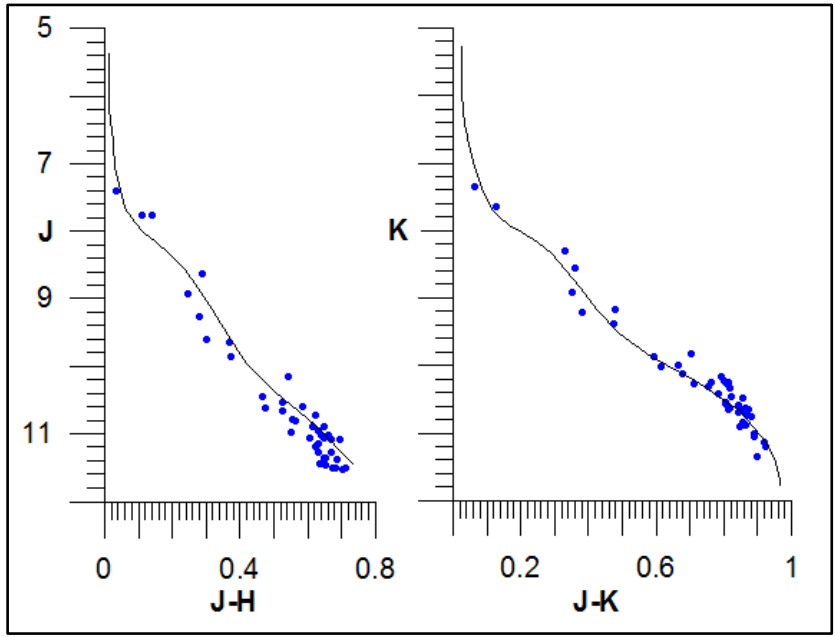

Fig 6: The membership validity of some variable stars in Pleiades' area; lying in the same CMDs of the cluster.

Table 2: The photometric data which obtained for the 63 variables of Taurus and located in Pleiades.

| # | Name | r | RA J2000 | DE J2000 | J | H | K | J-H | J-K |
|---|------|---|----------|----------|---|---|---|-----|-----|
| 1 | BU | 34.3851 | 57.2978 | 24.5194 | 14.093 | 13.626 | 13.552 | 0.467 | 0.541 |
| 2 | NT | 49.4368 | 56.1516 | 23.5965 | 15.894 | 15.241 | 14.795 | 0.653 | 1.099 |
| 3 | NY | 37.008 | 56.244 | 24.3912 | 14.711 | 14.318 | 14.126 | 0.393 | 0.585 |
| 4 | NZ | 30.9975 | 56.3047 | 24.2566 | 14.032 | 13.405 | 13.195 | 0.627 | 0.837 |
| 5 | OT | 33.7789 | 56.3732 | 23.7605 | 11.063 | 10.417 | 10.251 | 0.646 | 0.812 |
| 6 | OW | 26.881 | 56.4539 | 23.8529 | 11.476 | 10.822 | 10.661 | 0.654 | 0.815 |
| 7 | PP | 12.0526 | 56.6299 | 24.1174 | 11.284 | 10.614 | 10.461 | 0.67 | 0.823 |
| 8 | QS | 7.455 | 56.7959 | 24.231 | 14.269 | 13.72 | 13.398 | 0.549 | 0.871 |
| 9 | QX | 9.5706 | 56.9227 | 23.9719 | 11.066 | 10.461 | 10.312 | 0.605 | 0.754 |
| 10 | V0338 | 37.7473 | 56.9989 | 24.7313 | 10.952 | 10.321 | 10.159 | 0.631 | 0.793 |
| 11 | V0343 | 26.2268 | 57.072 | 24.5045 | 11.516 | 10.806 | 10.669 | 0.71 | 0.847 |
| 12 | V0345 | 38.2476 | 57.0858 | 23.5171 | 14.431 | 13.816 | 13.533 | 0.615 | 0.898 |
| 13 | V0352 | 17.0652 | 57.1478 | 24.2009 | 14.083 | 13.446 | 13.24 | 0.637 | 0.843 |
| 14 | V0357 | 14.7287 | 57.0471 | 23.9501 | 16.251 | 15.903 | 15.818 | 0.348 | 0.433 |
| 15 | V0360 | 42.2231 | 57.3258 | 23.564 | 13.665 | 13.146 | 13.108 | 0.519 | 0.557 |
| 16 | V0369 | 39.8139 | 57.5465 | 23.9284 | 14.467 | 13.877 | 13.725 | 0.59 | 0.742 |
| 17 | V0370 | 45.7666 | 57.5517 | 24.533 | 10.983 | 10.433 | 10.271 | 0.55 | 0.712 |
| 18 | V0448 | 14.1959 | 56.6496 | 23.967 | 11.513 | 10.832 | 10.652 | 0.681 | 0.861 |
| 19 | V0454 | 11.902 | 56.7759 | 24.3035 | 13.863 | 13.426 | 13.416 | 0.437 | 0.447 |
| 20 | V0460 | 17.4625 | 57.124 | 23.9684 | 13.88 | 13.229 | 13.001 | 0.651 | 0.879 |
| 21 | V0463 | 35.6185 | 57.189 | 24.624 | 14.078 | 13.465 | 13.172 | 0.613 | 0.906 |

Tadross, A. L., Hanna, M. A., Awadalla, N. S.

Table 2: continued.

| #  | Name  | r       | RA J2000 | DE J2000 | J      | H      | K      | J-H   | J-K   |
|----|-------|---------|----------|----------|--------|--------|--------|-------|-------|
| 22 | V0527 | 24.4804 | 56.5857  | 24.4463  | 14.486 | 14.186 | 14.173 | 0.3   | 0.313 |
| 23 | V0529 | 24.8685 | 56.6032  | 24.4651  | 13.651 | 13.266 | 13.145 | 0.385 | 0.506 |
| 24 | V0530 | 12.6831 | 56.6665  | 23.9881  | 14.431 | 14.132 | 13.986 | 0.299 | 0.445 |
| 25 | V0541 | 22.939  | 56.9216  | 23.7403  | 11.361 | 10.714 | 10.555 | 0.647 | 0.806 |
| 26 | V0545 | 21.2057 | 57.2055  | 23.9773  | 10.658 | 10.133 | 9.992  | 0.525 | 0.666 |
| 27 | V0554 | 44.8437 | 57.5563  | 23.7404  | 13.555 | 13.143 | 13.06  | 0.412 | 0.495 |
| 28 | V0633 | 47.9025 | 56.153   | 24.601   | 14.932 | 14.466 | 14.414 | 0.466 | 0.518 |
| 29 | V0646 | 5.0115  | 56.8256  | 24.0365  | 11.074 | 10.38  | 10.265 | 0.694 | 0.809 |
| 30 | V0647 | 1.6994  | 56.8306  | 24.1391  | 7.767  | 7.627  | 7.639  | 0.14  | 0.128 |
| 31 | V0650 | 26.3258 | 56.8618  | 23.6784  | 7.406  | 7.37   | 7.343  | 0.036 | 0.063 |
| 32 | V0651 | 37.1801 | 56.91    | 24.7343  | 14.457 | 14.051 | 13.96  | 0.406 | 0.497 |
| 33 | V0657 | 24.8378 | 57.0369  | 24.4943  | 15.331 | 15.144 | 14.782 | 0.187 | 0.549 |
| 34 | V0669 | 46.6573 | 57.507   | 24.6135  | 14.479 | 13.905 | 13.578 | 0.574 | 0.901 |
| 35 | V0703 | 34.1046 | 56.8903  | 24.6842  | 10.807 | 10.244 | 10.128 | 0.563 | 0.679 |
| 36 | V0758 | 26.5463 | 57.1479  | 23.7682  | 13.659 | 13.364 | 13.369 | 0.295 | 0.29  |
| 37 | V0787 | 22.9489 | 56.4337  | 24.0741  | 11.533 | 10.829 | 10.662 | 0.704 | 0.871 |
| 38 | V0788 | 35.7599 | 56.5054  | 23.6111  | 15.442 | 14.683 | 14.539 | 0.759 | 0.903 |
| 39 | V0789 | 31.9985 | 56.5287  | 24.5628  | 11.013 | 10.351 | 10.25  | 0.662 | 0.763 |
| 40 | V0793 | 11.7098 | 56.9182  | 24.302   | 11.147 | 10.517 | 10.328 | 0.63  | 0.819 |
| 41 | V0794 | 6.4847  | 56.9663  | 24.0965  | 13.8   | 13.207 | 13.023 | 0.593 | 0.777 |
| 42 | V0813 | 32.2802 | 56.5271  | 24.5674  | 10.793 | 10.236 | 10.113 | 0.557 | 0.68  |

Table 2: continued.

| # | Name | r | RA J2000 | DE J2000 | J | H | K | J-H | J-K |
|---|------|---|----------|----------|---|---|---|-----|-----|
| 43 | V0814 | 11.4522 | 56.6641 | 24.0297 | 10.454 | 9.986 | 9.859 | 0.468 | 0.595 |
| 44 | V0815 | 24.2796 | 56.8064 | 23.7143 | 10.624 | 10.148 | 10.009 | 0.476 | 0.615 |
| 45 | V0851 | 34.6691 | 56.2927 | 24.3922 | 14.363 | 13.64 | 13.386 | 0.723 | 0.977 |
| 46 | V0854 | 39.4209 | 56.3646 | 23.6326 | 14.163 | 13.531 | 13.269 | 0.632 | 0.894 |
| 47 | V0855 | 38.6823 | 56.4174 | 24.6272 | 8.624 | 8.337 | 8.294 | 0.287 | 0.33 |
| 48 | V1006 | 51.8987 | 56.2914 | 23.4192 | 13.704 | 13.214 | 13.137 | 0.49 | 0.567 |
| 49 | V1007 | 46.6558 | 56.3587 | 24.753 | 14.914 | 14.631 | 14.64 | 0.283 | 0.274 |
| 50 | V1010 | 25.8624 | 56.4604 | 23.874 | 11.02 | 10.382 | 10.219 | 0.638 | 0.801 |
| 51 | V1016 | 31.9607 | 57.1803 | 24.5565 | 15.001 | 14.594 | 14.521 | 0.407 | 0.48 |
| 52 | V1044 | 16.9547 | 56.5579 | 24.0237 | 14.288 | 13.915 | 13.782 | 0.373 | 0.506 |
| 53 | V1045 | 30.5685 | 56.5945 | 24.5702 | 9.274 | 8.994 | 8.923 | 0.28 | 0.351 |
| 54 | V1065 | 46.4357 | 56.5172 | 23.4055 | 10.54 | 10.014 | 9.836 | 0.526 | 0.704 |
| 55 | V1085 | 24.8671 | 56.3975 | 24.0832 | 8.921 | 8.673 | 8.562 | 0.248 | 0.359 |
| 56 | V1088 | 37.5006 | 57.3371 | 24.5571 | 14.796 | 14.272 | 14.125 | 0.524 | 0.671 |
| 57 | V1089 | 32.0883 | 57.3502 | 23.8393 | 9.854 | 9.482 | 9.378 | 0.372 | 0.476 |
| 58 | V1090 | 35.249 | 57.388 | 23.7954 | 9.595 | 9.294 | 9.214 | 0.301 | 0.381 |
| 59 | V1171 | 22.8488 | 56.6184 | 24.4339 | 9.638 | 9.27 | 9.159 | 0.368 | 0.479 |
| 60 | V1172 | 39.0658 | 56.9083 | 23.4681 | 11.192 | 10.567 | 10.41 | 0.625 | 0.782 |
| 61 | V1173 | 36.7509 | 57.5211 | 24.124 | 11.366 | 10.715 | 10.56 | 0.651 | 0.806 |
| 62 | V1189 | 16.665 | 56.5536 | 24.0544 | 11.435 | 10.798 | 10.594 | 0.637 | 0.841 |
| 63 | V1210 | 8.5927 | 56.7676 | 23.9952 | 7.77 | 7.658 | 7.642 | 0.112 | 0.128 |